%% file: main.tex
\documentclass[manuscript]{acmart}

\usepackage[utf8]{inputenc} %
\usepackage[T1]{fontenc}    %
\usepackage{hyperref}       %
\usepackage{url}            %
\usepackage{booktabs}       %
\usepackage{amsfonts}       %
\usepackage{nicefrac}       %
\usepackage{microtype}      %
\usepackage{xcolor}         %
\usepackage{multirow}
\usepackage{graphicx}
\usepackage{tabularx}
\usepackage{cleveref}
\usepackage{subcaption}

\usepackage[compact]{titlesec}
\titlespacing{\subsection}{0pt}{1.1ex}{0ex}

\title{Design Principles of Robust Multi-Armed Bandit Framework in Video Recommendations}

\author{Belhassen Bayar}
\authornote{Both authors contributed equally to this research \\ Copyright held by the author(s).}
\email{bayarb@amazon.com}
\author{Phanideep Gampa}
\authornotemark[1]
\email{phanide@amazon.com}

\affiliation{%
  \institution{Amazon Prime Video}
  \country{United States} 
}

\author{Ainur Yessenalina}
\email{yessenal@amazon.com}
\author{Zhen Wen}
\email{zhenwen@amazon.com}
\affiliation{%
  \institution{Amazon Prime Video}
  \country{United States} 
  }

\begin{document}

\begin{abstract}
\input{Sections_Belhassen/Abstract}
\end{abstract}


	
	\maketitle

\input{Sections_Belhassen/Intro_v0}
	
\input{Sections_Belhassen/Related_Work_v0}

\input{Sections_Belhassen/CurrentSystem}

\input{Sections_Belhassen/Approach_v0}

\input{Sections_Belhassen/Experiments_v0}

\input{Sections_Belhassen/Conclusion_v0}

	\bibliographystyle{plain}
	\bibliography{main.bib}
	
	\newpage

\end{document}

%% file: Sections_Belhassen/Abstract.tex
Current multi-armed bandit approaches in recommender systems (RS) have focused more on devising effective exploration techniques, while not adequately addressing common exploitation challenges related to distributional changes and item cannibalization. Little work exists to guide the design of robust bandit frameworks that can address these frequent challenges in RS. In this paper, we propose a new design principles to (i) make bandit models robust to time-variant metadata signals, (ii) less prone to item cannibalization, and (iii) prevent their weights fluctuating due to data sparsity.
Through a series of experiments, we systematically examine the influence of several important bandit design choices. We demonstrate the advantage of our proposed design principles at making bandit models robust to dynamic behavioral  changes through in-depth analyses. Noticeably, we show improved relative gain compared to a baseline bandit model not incorporating our design choices of up to $11.88\%$ and $44.85\%$, respectively in ROC-AUC and PR-AUC. Case studies about fairness in recommending specific popular and unpopular titles are presented, to demonstrate the robustness of our proposed design at addressing popularity biases.

%% file: Sections_Belhassen/Intro_v0.tex
\section{Introduction}
Capturing users’ interests that change over time is a key point in personalized video recommendations. This is because users' interests in recommender systems (RS) are typically time and catalog dependent, which can potentially make recommendations stale. This becomes very challenging when the promotional basket is organically updated with a set of entering and exiting titles; each associated with different degrees of popularity and release recency that evolve over time. 
In order to address recommendation staleness in RS with respect to both user's and title's behavioral changes, researchers developed numerous algorithms to solve exploration/exploitation in multi-armed bandit problems~\cite{bouneffouf2020survey}. In video recommendation settings, the objective is to select for a user the arm (title) with the highest expected reward while exploring other arms to gain information about their expected rewards. Contextual bandit models are cheap to train, and have shown promising capabilities 
in a continual incremental training setting~\cite{bouneffouf2020survey,suhr2022continual,cavenaghi2021non,silva2022multi}.

While most of the bandit research in RS has focused on developing exploration techniques to address several challenges associated with recency and distributional changes~\cite{chu2011contextual,chapelle2011empirical, zhou2015survey}, there is little research work that specifically focused on addressing exploitation related bandit issues. However, exploitation issues are very common in RS in general, including bandit models, such as distributional and categorical biases induced by data sparsity or lack of granular representation of both title and user entities~\cite{wang2019enhancing, han2019adaptive,abdollahpouri2019unfairness,mougan2022fairness}. Specifically, because of their small size and linear nature, linear bandit model parameters in continual incremental training setting can be associated with abrupt and non-smooth changes along with biases towards recommending more popular titles or categories defined through the input contextual signals. Additionally, many behavioral and time-variant metadata signals in RS can change dramatically during title promotion initial phase, creating distributional discrepancy in the user and/or title representation between training and inference phases which can drastically impact a bandit model performance.

In this paper, we present our design principles to tackle common exploitation challenges in building a multi-armed bandit production framework for video recommendations. We systematically investigate three common known issues in bandit models, and RS in general while proposing a solution for each of the challenges: (1) A novel data augmentation (DA) solution in RS inspired by computer vision research~\cite{shorten2019survey} that can help the bandit model to generalize to different distributional changes of time-variant metadata such as title recency. DA is also designed to address sparsity issues associated with this category of time-variant metadata.~(2) Temporal title signals that can model an arm performance over consecutive time slots to overcome cannibalization issues due to lack of granular representation of an arm when multiple titles share similar contextual metadata information. This type of signal can also help the bandit model to quickly adapt to the very recent changes in the viewership trends.~(3) $L2$ smoothing of bandit's parameter weights to attenuate their fluctuations and abrupt changes due to data sparsity in a continual incremental training setting. Our smoothing method incorporates an adaptive regularization techniques where the smoothing parameters are learned from the bandit historical weights, and is designed to prevent popularity biases towards a title or a category of titles.

Through a set of experimental results, we demonstrate the advantage of our proposed design principles to make a multi-armed bandit framework robust to common dynamic challenges encountered in RS; and particularly more frequent in the context of promotional catalogue with cold start titles. We thoroughly evaluate our proposed design choices through traditional AUC scores, along with a customized nDCG (normalized discounted cumulative gain) ranking metrics that can be leveraged in any bandit offline setting. The new designed framework typically outperforms a baseline bandit model used in production over 40 consecutive offline runs. Noticeably, the proposed model can achieve $11.88\%$ and $44.85\%$ relative gain compared to baseline, respectively in ROC-AUC and PR-AUC.

\begin{figure*}[ht]

    \begin{subfigure}{1\textwidth}
    \centering
    \includegraphics[scale=0.15]{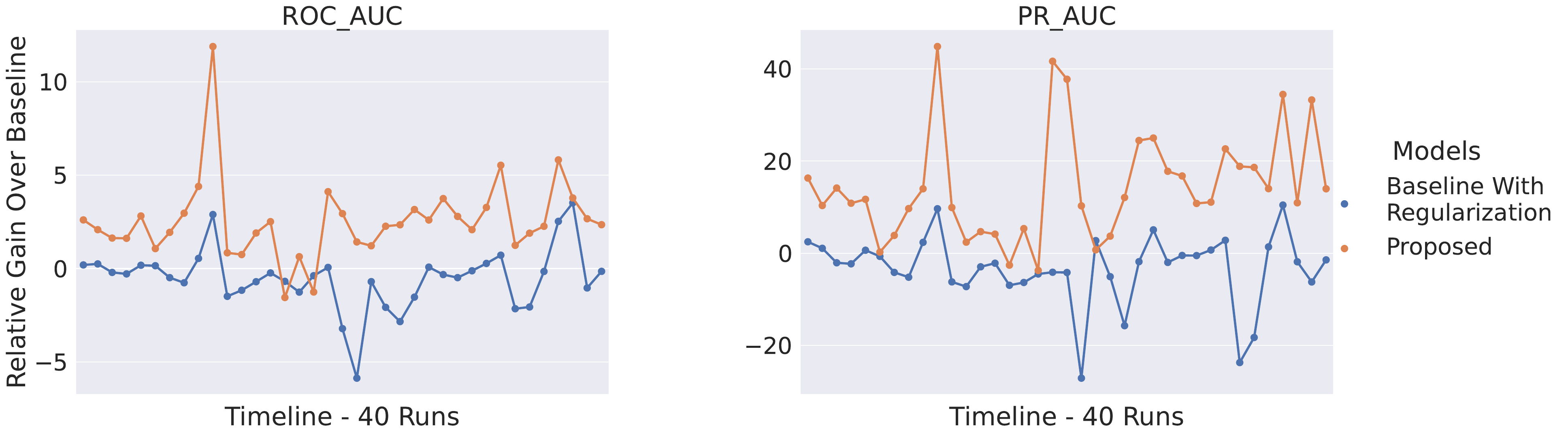}
    \caption{Proposed model vs Baseline model with Regularization}
    \label{fig:auc_1}
    \end{subfigure}
    \begin{subfigure}{1\textwidth}
    \centering
    \includegraphics[scale=0.15]{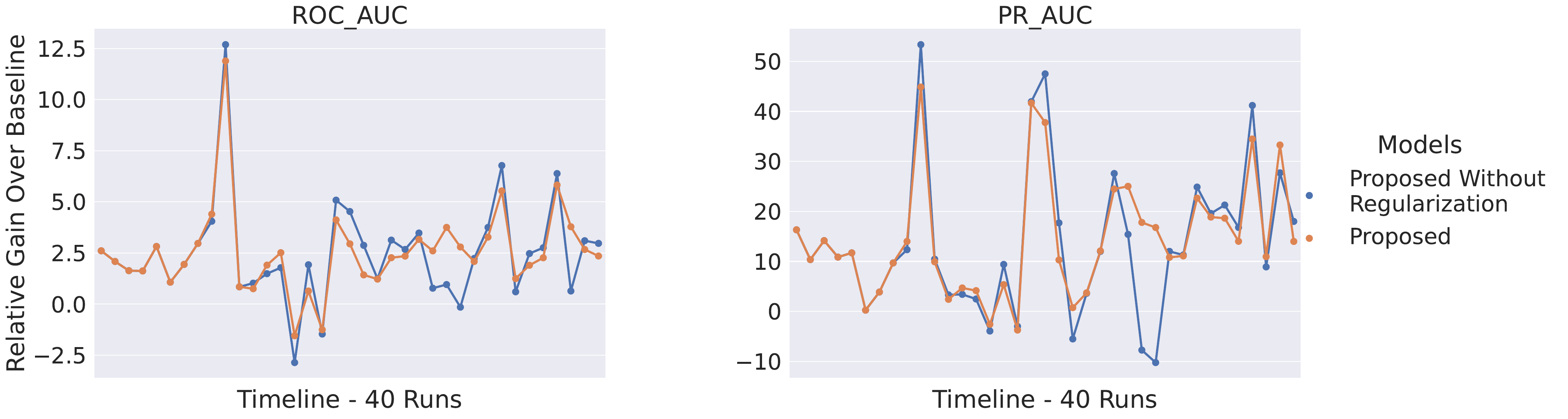}
    \caption{Proposed model vs Proposed model without Regularization}
    \label{fig:auc_2}
    \end{subfigure}
    
     \begin{subfigure}{1\textwidth}
    \centering
    \includegraphics[scale=0.15]{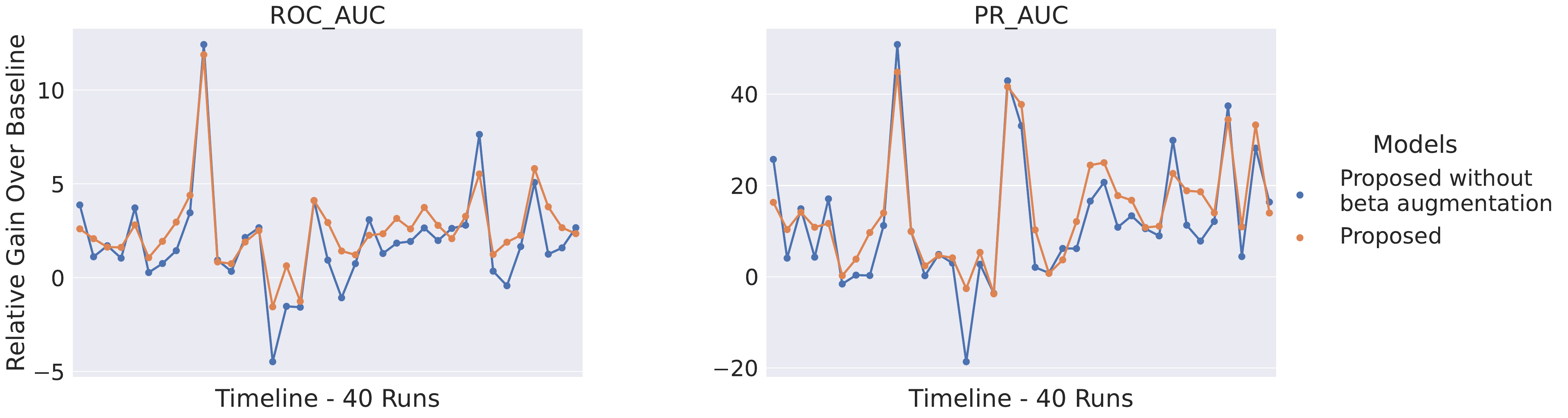}
    \caption{Proposed model vs Proposed model without the $beta$ augmentation step}
    \label{fig:auc_3}
    \end{subfigure}
    \caption{The above figures contain the relative AUC scores gain of various model configurations over the baseline model. The relative gain is calculated for 40 consecutive training runs. We can observe that the proposed techniques help the model achieve better performance consistently.}
    \label{fig: auc plots}
\end{figure*}

%% file: Sections_Belhassen/Related_Work_v0.tex
\section{Related Work}
The existing contextual bandit frameworks balance exploration/exploitation trade offs either by updating the confidence intervals in UCB based algorithms~\cite{chu2011contextual} or by updating the priors in Thompson sampling based algorithms~\cite{chapelle2011empirical, zhou2015survey}. While UCB based algorithms use confidence intervals for exploring the arms, Thompson sampling based bandits use variance of the prior distributions to control the exploration. Also, collaborative filtering based algorithms use side information for exploiting and enriching the user and item information to deal with sparse data~\cite{zheng2017joint, wang2019enhancing, abdollahpouri2019unfairness, han2019adaptive}. Furthermore, using one hot encoding can result in group category based biases where positive rewards of one group within the same category can drive the value of the category weight~\cite{mougan2022fairness}. Target encoding based techniques that use continuous values for encoding can help prevent categorical sparsity biases~\cite{pargent2019benchmark, mougan2022fairness}. Tree based models use heuristics like distribution based imputation for adding pseudo instances to deal with various categorical sparsity and biases~\cite{au2018random}. Additionally, adapting machine learning based models to various invariances in the input data distributions through data augmentation techniques is a well studied problem in computer vision~\cite{shorten2019survey}. Also, data augmentation by corrupting the item sequences to deal with data sparsity helps improve the performance of sequential recommender systems~\cite{song2022data}.

%% file: Sections_Belhassen/CurrentSystem.tex
\section{Current Bandit Ranker}
\label{sec:Current system}
The current model is a logistic regression based Bayesian bandit with thompson sampling~\cite{chapelle2011empirical, zhou2015survey} where each arm represents a title. We start with a random prior for arm feature weights that gets updated during training. The model takes input a set of features corresponding to the arms such as one hot encoded signals, e.g., Marketing Classification, content category, and Hours lapsed since the launch of the title. 
The model is trained to predict the user streaming action as a positive reward within a time-based attribution window, given a user-title pair. During the model training, we employ a binary cross entropy loss~\cite{chapelle2011empirical, bishop2006pattern}, and the mean weights are shared across the arms as we try to rank the titles among themselves at any given time. The covariance of features is tracked separately for each title. During inference, given a customer and set of titles the model emits scores for each title which are used to rank the titles. Similar to all online models, this model is incrementally trained every T hours. The set of arms is updated every time a new title enters or exits the recommendation catalogue. Refer to the section~\ref{sec:Experiments} for more details.

\subsection{Challenges}
\label{subsec: challenges}
Because of their online and linear nature, bandit based recommendation systems commonly have to deal with 1) categorical data sparsity where the number of training examples for certain categories are very less; 2) categorical cannibilisation due to the lack of granular metadata; and 3) time based feature columns that dynamically update the input representations. These challenges are common in any ML models in RecSys and particularly in exploitation. 

\textbf{Hours lapsed since the launch of the title (recency)} This feature helps to factor in the recency of the title launch and to tune the exploration/exploitation trade-offs accordingly. The feature value is set based on a time based binning of first $hr_1$ hours, $hr_1$-$hr_2$ hours, $hr_2$-$hr_3$ hours and greater than $hr_3$ hours respectively. While other features used by the model are fixed given a title-user pair, this feature can change over time. Because of the dynamic nature of this feature, the performance of a title moving from one bin to the next bin with time could be affected by the new bin weight. A high performing title in old bin can be down ranked when it moves to new bin if the weight of the new bin is less when compared with other time based bins, and vice versa with low performing title being up-ranked. 

\textbf{Categorical Feature Biases and Sparsity} Apart from the dynamic time based feature, the model takes fixed one hot encoded categorical features as the input. Such encoding comes with a bias where a high performing title from one category will drive the performance of other titles from the same category by giving higher scores than expected, and vice versa. At the same time, one hot encoded features suffer with data sparsity in the context of online models where the set of titles and their categories can change. This data sparsity results in abrupt fluctuation of weights and can impact the next incremental training.

\begin{figure*}[ht]
    
    \begin{subfigure}{1\textwidth}
    \centering
    \includegraphics[scale=0.15]{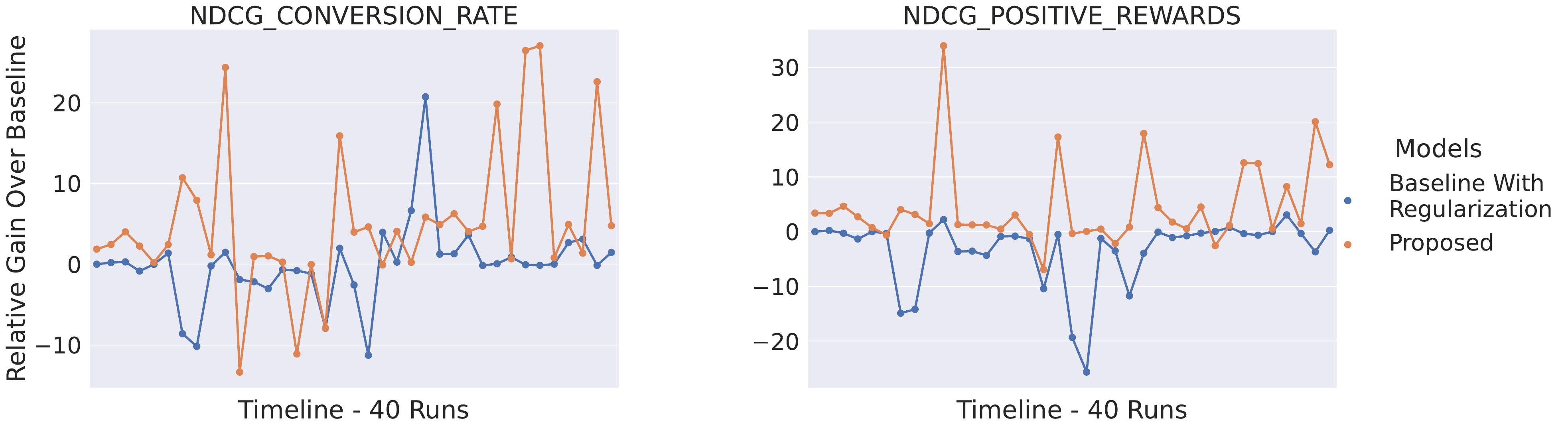}
    \caption{Proposed model vs Baseline model with Regularization}
    \label{fig:ndcg_1}
    \end{subfigure}
    \begin{subfigure}{1\textwidth}
    \centering
    \includegraphics[scale=0.15]{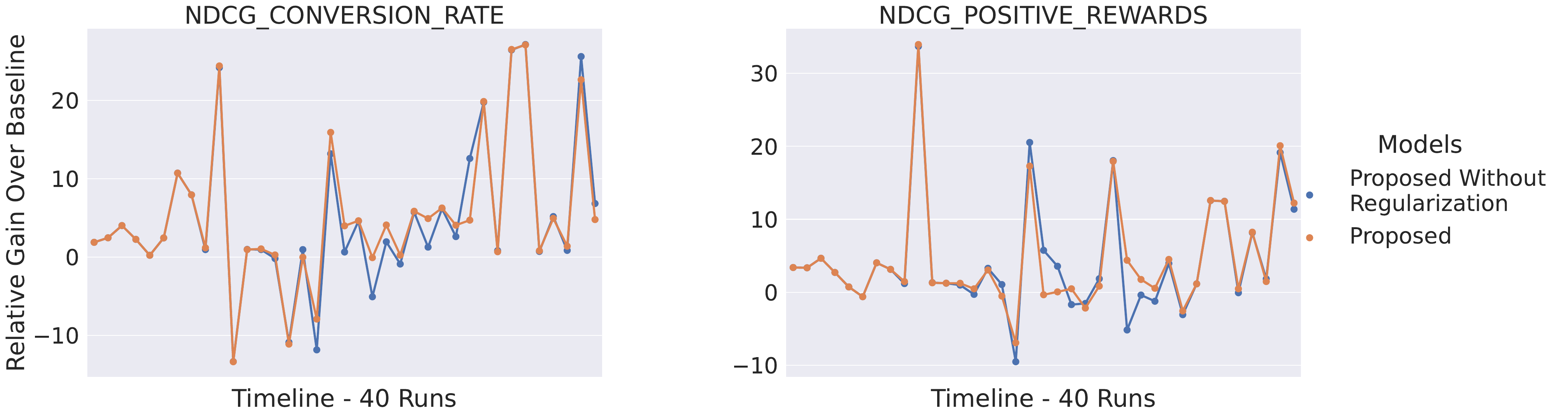}
    \caption{Proposed model vs Proposed model without Regularization}
    \label{fig:ndcg_2}
    \end{subfigure}
    
     \begin{subfigure}{1\textwidth}
    \centering
    \includegraphics[scale=0.15]{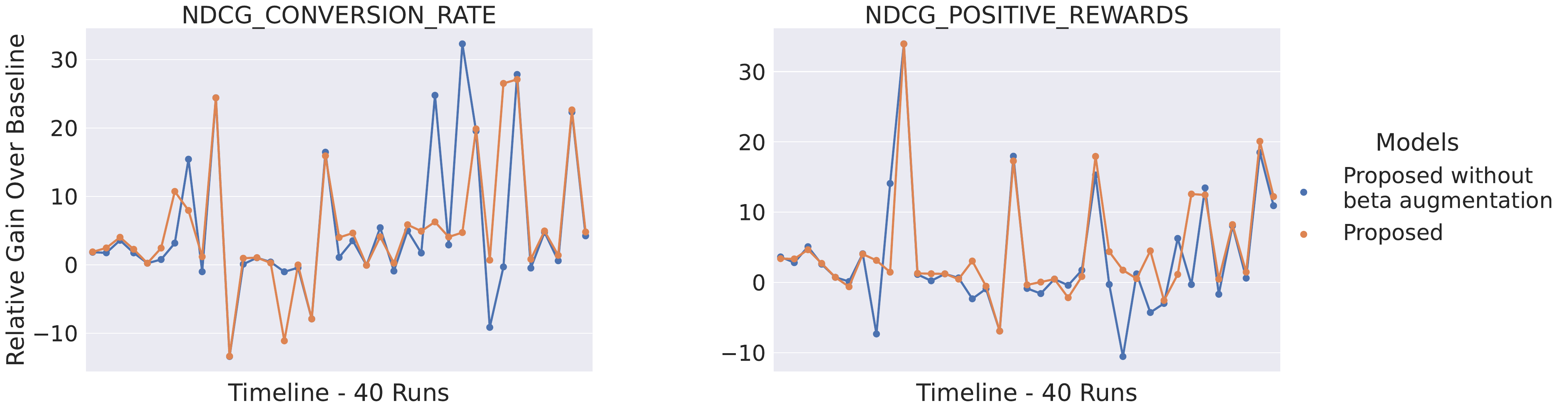}
    \caption{Proposed model vs Proposed model without the $\beta$ augmentation step}
    \label{fig:ndcg_3}
    \end{subfigure}
    \caption{The above figures contain the relative nDCG metrics gain of various model configurations over the baseline model. As defined in section~\ref{subsec: results}, the aggregated nDCG metrics quantify the alignment between predicted title SOVs and production time title conversion rates and title positive rewards. The relative gain is calculated for 40 consecutive training runs.}
    \label{fig: ndcg plots}
\end{figure*}

%% file: Sections_Belhassen/Approach_v0.tex
\section{Approach}

\label{sec:Approach} 

To build a robust multi armed bandit framework by efficiently dealing with dynamic changes in recency features and the categorical feature biases, we propose the following design principles.

\textbf{Data Augmentation (DA)} This approach is designed to fix  fluctuations and feature mismatch sparsity issues of the hours since launch bin feature. As part of this approach, we perform a two stage data augmentation process.  Specifically, 1) for each title, we create a copy of a small $\alpha\%$ ratio from its training examples where we transition the recency bin to the next bin. Titles that belong to the last bin will remain in the same bin in the new $\alpha\%$ sampled data; and 2) even after the first stage of title level sampling, there can be sparsity in some bins, we similarly upsample them with randomly selected training examples to guarantee at least $\beta\%$ shares with respect to the overall bin distribution. If there are zero examples from a particular hours since launch bin during training, the previous biases present in the weights would propagate to the next title that enters into that bin. The $\beta\%$ upsampling step helps attenuate this problem by providing non zero feedback from all the titles as an average. During training phase, we exclude the $\beta\%$ randomly upsampled data from updating the variance vectors of the title arms in order to prevent additional noise influencing the exploration. For our experiments, we decided to use $\alpha = 15\%$ and $\beta = 10\%$ based on empirical study of various configurations.

\textbf{Temporal Title Signals (TS)} This approach is aimed at tackling the categorical bin level cannibalization by providing granular features that can differentiate at title/arm level. At the same time, these signals incentivise exploitation by factoring in the relative popularity of the titles as input features. In this approach, we add three new columns containing title level Normalised Distinct Stream (NDS) values calculated in the last three $p$ hour slots. Specifically, following the definition given in equation~\ref{eq: nds}, we calculate $NDS_{title, (T,    T-p)}, NDS_{title, (T-(p+1), T-2p)}, NDS_{title, (T-(2p + 1), T-3p)}$ values for each title during the run hour $T$. The three new columns encode the popularity of each title relative to other titles in the three different time slots. A title would be up-ranked if it consistently has high NDS in all three slots and vice versa. For cold start titles that may not have data for all the three columns, we use average NDS observed during training for the respective columns. For cold start titles with high priority marketing classification, we give train time maximum NDS value for all columns during their first $hr_1$ hours after launch. We observed that this strategy boosts share of voice (SOV) of such titles without impacting the performance of other titles. Share of voice of a title is defined as the percentage of users for which this title is present in top position when the titles are sorted by model scores. 

\begin{equation}
	\label{eq: nds}
	NDS_{title, time-period} = \frac{distinct-streams(title,time-period)}{\sum_{t\in title-set} distinct-streams(t,time-period)} 
\end{equation}

\textbf{L2 smoothing of weights} To prevent the fluctuation of weights due to the data sparsity of various categorical features, we regularize the weights of the model by taking historical average of the weights as reference for smoothing. Specifically, we apply L2 smoothing for feature columns corresponding to marketing classification, content category, hours since launch bin, and the newly added NDS features by adding the L2 cost defined in equation~\ref{eq: L2 smoothing 2} during training. As part of this approach, we use a smoothing vector as a reference to control the movements of weights and attenuate sudden spikes. As given in equation~\ref{eq: L2 smoothing}, we use the average weights of the previous $q$ train runs as the smoothed weight vector for the regularization process. Given a regularization parameter $\lambda$, the L2 smoothing loss at training run hour T is defined as:

\begin{align}
	\label{eq: L2 smoothing}
	W_{T, smooth} &= \frac{\sum_{t=1}^{t=q} W_{T-t}}{q}  \\
	\label{eq: L2 smoothing 2}
	L2-smoothing-loss(W_{T}) &= \lambda||W_{T}-W_{T, smooth}||_2
\end{align}

%% file: Sections_Belhassen/Experiments_v0.tex
\section{Experiments}

\label{sec:Experiments}

\subsection{Data}
For the experiments, the training and validation data is created by down sampling the production logs to maintain an X number of negatives for each positive user and title pair. This data is curated every T hours from the prod logs containing user request ids and the recorded streaming actions. For evaluation, we take a down sampled version of the prod logs gathered every T hours following the cadence for training. The production data consists of titles in the order of 50 and several dozens of millions of users. At any given time, the set of titles are updated based on a title's promotional window.

\subsection{Experiment Setup}

Thompson sampling based Bayesian bandit model~\cite{chapelle2011empirical} consists of one linear layer with a sigmoid activation function, where Bayesian priors are maintained per feature and arm/title. As described in section~\ref{sec:Current system}, the mean weights are shared across the arms, but variance is kept separate.%
 The proposed model takes the 3 additional temporal signal (TS) features compared to the baseline. %
For training, we use Adam~\cite{kingma2014adam} optimizer, and the models are trained end-to-end for 3 epochs using the binary cross entropy loss. We also apply a custom regularization term on the Bayesian priors.
The model is incrementally trained on data gathered every T hours. For robust evaluation, we perform incremental training and evaluation for 40 consecutive runs on the data gathered from the production. For L2 regularization, we use a $0.25$ regularization constant.

\subsection{Results}

\label{subsec: results}

We evaluate the models for stream action prediction given a user-title pair by calculating ROC-AUC and PR-AUC scores for each training run. We also report two aggregated nDCG (normalized discounted cumulative gain) based metrics that are useful for determining whether the title share of voice (SOV) re-adjustments are happening in the right direction when compared with production logs. Specifically, we report the nDCG metric calculated between the ranked lists of predicted top1 sov and the production time conversion rates (nDCG Conversion Rates), positive rewards (nDCG Positive Rewards) for all the titles. So, a higher nDCG metric and a drop in top1 SOV for a specific title implies that the re-adjustments are happening in the right direction at an aggregate level. For all the metrics reported, we provide the relative gain of the proposed model and its configurations over the baseline model. We compare the metrics of the below model configurations against the baseline model and report the relative gain over baseline: 1) Proposed: This refers to the proposed model where we apply 3 methods described in section~\ref{sec:Approach}; 2) Baseline With Regularization: This is the baseline model but with regularization applied to categorical weights; 3) Proposed without Regularization: This is the proposed model without the regularization loss; 4) Proposed without $beta$: This is the proposed model, but without the second level of augmentation using the $beta$ ratio.

As shown in the figures~\labelcref{fig: auc plots,fig: ndcg plots}, we can observe that the proposed model is consistently doing better than the baseline with respect to both the AUC scores and nDCG metrics. We observe similar patterns in both the metrics across various configurations. This implies that the proposed techniques not only help in improving the title relevancy prediction performance, they also re-adjust the title SOVs in the right direction. From figures~\labelcref{fig:auc_1,fig:ndcg_1}, we can understand that using regularization for baseline doesn't necessarily help. Also from figures~\labelcref{fig:auc_2,fig:ndcg_2}, proposed model without regularization has slight drops in some of the runs signifying the positive impact of smoothing. From the figures~\labelcref{fig:auc_3,fig:ndcg_3}, it is evident that the second level of augmentation using $\beta$ parameter is helping the model to better deal with the sparsity issues of the time based feature.

\begin{figure*}[ht]
    
    \begin{subfigure}{1\textwidth}
    \centering
    \includegraphics[scale=0.15]{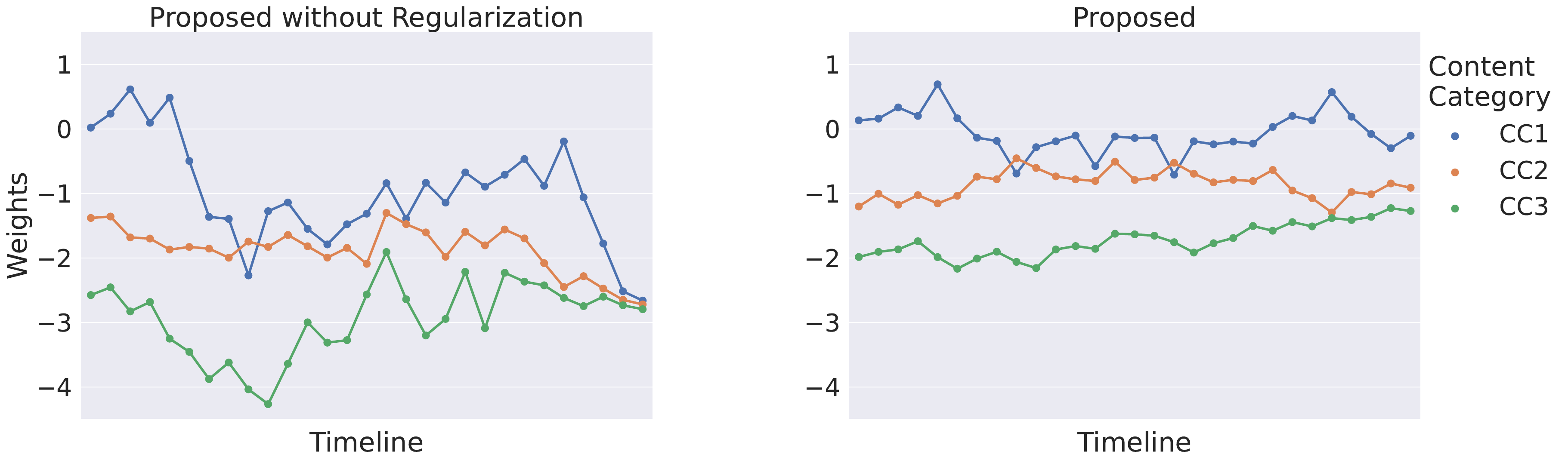}
    \caption{Content category feature weights}
    \label{fig:l2_content}
    \end{subfigure}
    \begin{subfigure}{1\textwidth}
    \centering
    \includegraphics[scale=0.15]{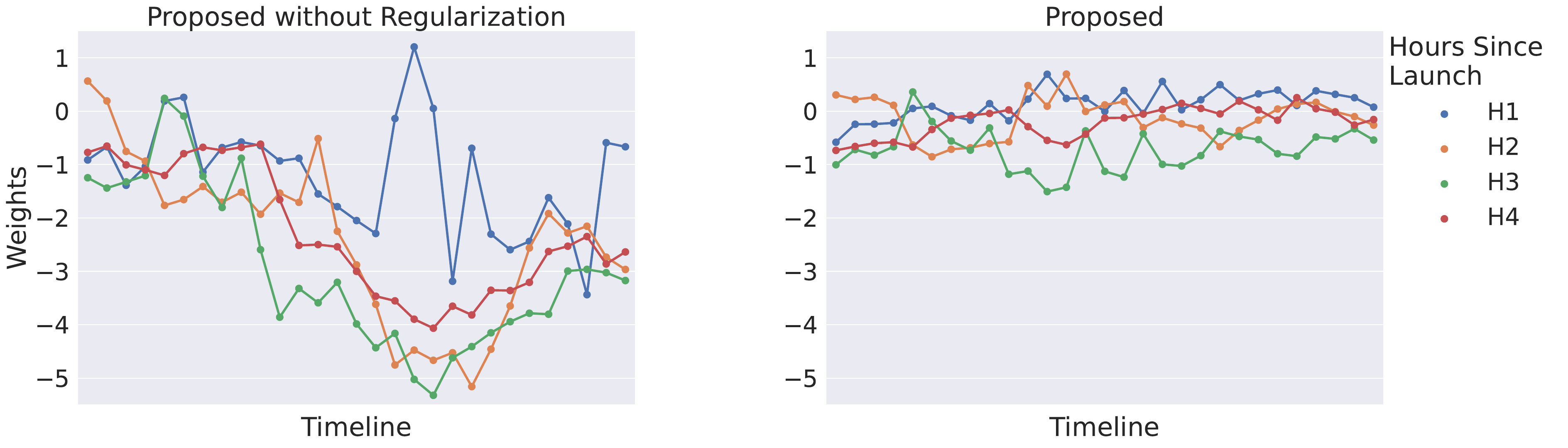}
    \caption{Hours since launch feature weights}
    \label{fig:l2_hours_since}
    \end{subfigure}
    
     \begin{subfigure}{1\textwidth}
    \centering
    \includegraphics[scale=0.15]{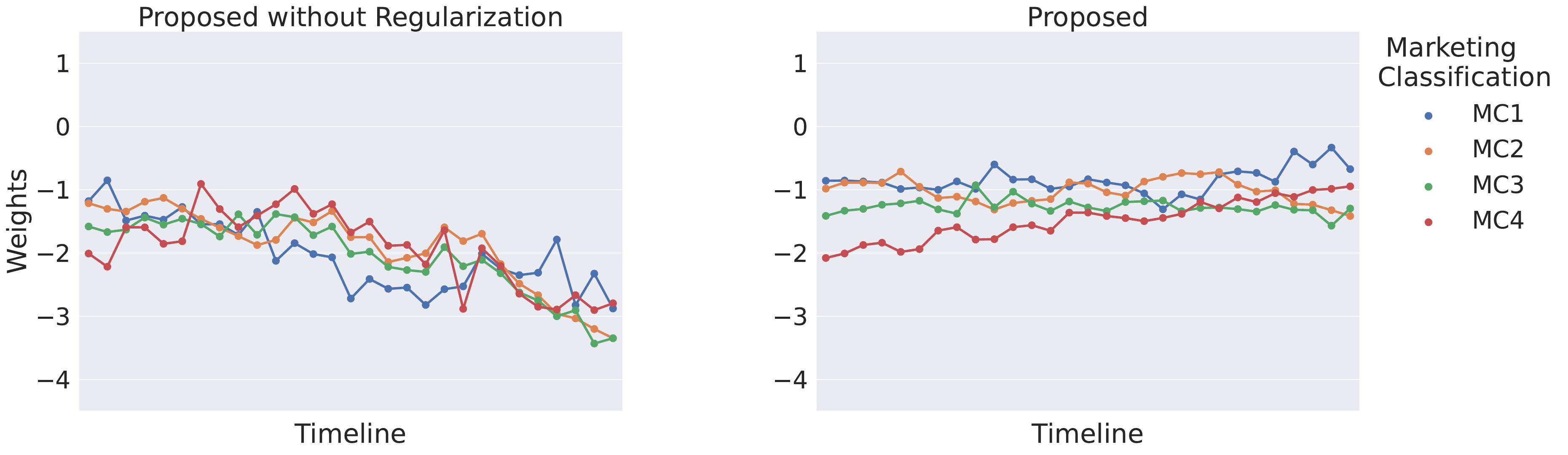}
    \caption{Marketing classification feature weights}
    \label{fig:l2_marketing}
    \end{subfigure}
     
    \begin{subfigure}{1\textwidth}
    \centering
    \includegraphics[scale=0.15]{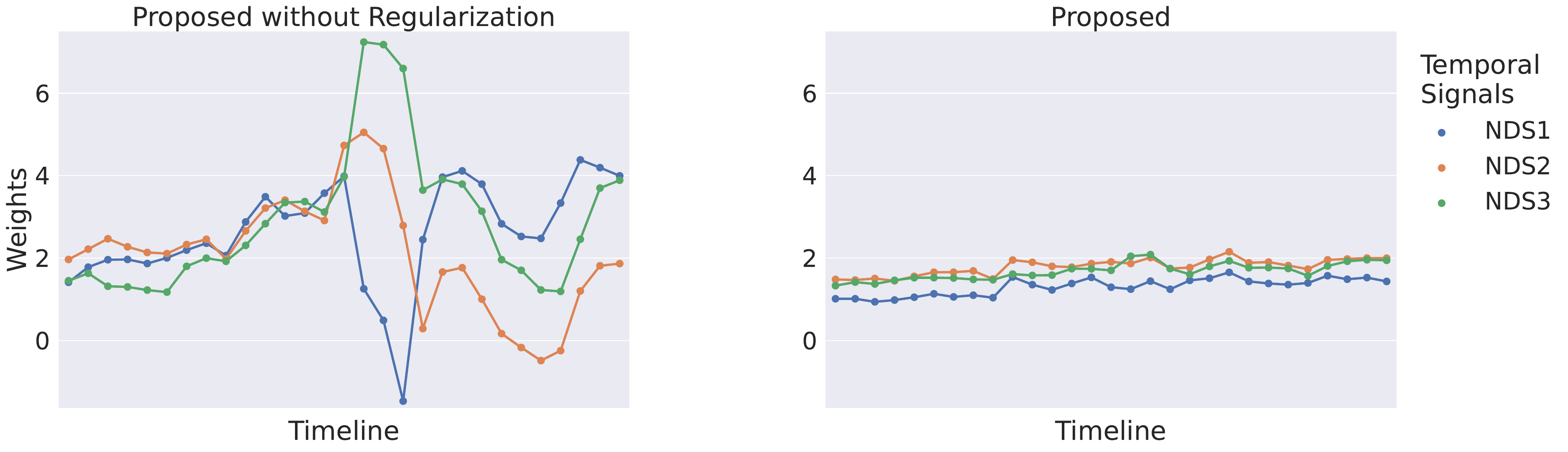}
    \caption{Temporal signal feature weights}
    \label{fig:l2_nds}
    \end{subfigure}
     
    \caption{Model feature weights before and after L2 regularization with a smoothing vector averaged over previous $q$ runs.}
    \label{fig:l2_weights}
\end{figure*}

\subsection{Case Studies: SOV allocation for specific titles}
The biases described in the section~\ref{subsec: challenges} can impact the SOV allocation of titles in an undesirable way. For example, an high performing title can get less SOV when it moves to the next recency bin that can be sparse at that point of time. At the same time, a low performing title can get more SOV than required when it belongs to a same category as another high performing title. As part of this case study, we analyse two such cases observed with prod model and how the proposed model can attenuate the biases by assigning correct SOV. As part of this case study, we report SOV of a title in its first position and also average rank across all customers. We report these metrics as a relative gain over baseline for two titles, a popular title for reference and also the title under study. 

\textbf{Over Exposure} As given in table~\ref{table: Over Exposure results}, the baseline model is giving more SOV to an over exposed title at the expense of a popular high performing title. This scenario was when a popular title moved to its next recency bin that was sparse. Observe that during the same runs, the proposed model with the help of data augmentation and temporal signals not only maintains the SOV of over exposed title, it still gives high SOV to the popular title than the baseline. 

 \begin{table}[h]
\caption{Average Rank and Top1 SOV relative gain of proposed model over the baseline for a popular title and overexposed title for three consecutive runs. For average rank, negative gain implies higher rank and for Top1 SOV, positive gain implies higher SOV than baseline.}
\label{table: Over Exposure results}
\begin{tabular}{ccccccc}
\toprule
&\multicolumn{2}{c}{Run1}&\multicolumn{2}{c}{Run2}&\multicolumn{2}{c}{Run3}\\
\cmidrule{2-3} \cmidrule{4-5}\cmidrule{6-7}
Titles & Avg Rank & Top1 SOV & Avg Rank & Top1 SOV & Avg Rank & Top1 SOV \\
\midrule

Popular Title &-43.41\%&19.73\%&-57.86\%&149.37\%&-22.56\%&32.51\% \\
Over Exposed Title &-18.34\%&21.72\%&-25.77\%&41.88\%&-13.80\%&-3.82\% \\

\bottomrule
\end{tabular}
\end{table}

\textbf{Under Exposure} As given in table~\ref{table: Under Exposure results}, the baseline model is giving significantly less SOV than expected to an under exposed title. This scenario is when the title belonged to a category with other low performing titles driving the performance. Observe that during the same runs, the proposed model with the help of temporal signals not only maintains the SOV of popular title, it gives significantly more SOV to the under exposed title than the baseline. 

 \begin{table}[h]
\caption{Average Rank and Top1 SOV relative gain of proposed model over the baseline for a popular title and a underexposed title for three consecutive runs. For average rank, negative gain implies higher rank and for Top1 SOV, positive gain implies higher SOV than baseline.}
\label{table: Under Exposure results}
\begin{tabular}{ccccccc}
\toprule
&\multicolumn{2}{c}{Run1}&\multicolumn{2}{c}{Run2}&\multicolumn{2}{c}{Run3}\\
\cmidrule{2-3} \cmidrule{4-5}\cmidrule{6-7}
Titles & Avg Rank & Top1 SOV & Avg Rank & Top1 SOV & Avg Rank & Top1 SOV \\
\midrule

Popular Title &-13.32\%&10.41\%&-56.62\%&200.43\%&-21.80\%&50.82\% \\
Under Exposed Title &1.59\%&-30.44\%&-29.06\%&710.05\%&-63.32\%&4871.34\% \\

\bottomrule
\end{tabular}
\end{table}

\subsection{Regularization of weights}
As given in the figure~\ref{fig:l2_weights}, regularization not only constrains the weights of the input features in a smaller region, it also reduces the fluctuations of weights over the training runs. While the smoothing of weights is clearly evident in non dynamic features like content category, marketing classification and temporal signals as given in figures~\labelcref{fig:l2_content,fig:l2_marketing,fig:l2_nds}. For dynamic features that change over time like hours since launch, the fluctuations can still be seen as evident in figure~\ref{fig:l2_hours_since}.

%% file: Sections_Belhassen/Conclusion_v0.tex
\section{Conclusion}
In this paper, we presented the primary steps to design a
robust multi-armed bandit production framework in video recommendations with high-value marketing content. We showed that data augmentation method, bandit weights regularization, along with temporal granular signals can address most common biases and title cannibalization challenges in RS. Through our experiments and case studies, we demonstrated the effect of using each of our design choices. Experimental results from offline analyses showed %
~ that our proposed method can achieve up to $11.88\%$ and $44.85\%$ relative gain compared to a bandit baseline model used in production, respectively in ROC-AUC and PR-AUC.